\documentclass[amsmath,amssymb,aps,prl,onecolumn,groupedaddress]{revtex4-1}
\usepackage{graphicx}
\allowdisplaybreaks
\newcommand{\mi}{\mathrm i}
\newcommand{\e}{\mathrm e}
\begin{document}


\title{Non-Markovian dynamics of open quantum systems without rotating wave approximation}


\author{Ming-Jia Tang$^{1}$, Yu-Kai Wu$^{1}$, Ming Lyu$^{1}$, Jia-Nan Tang$^{1}$, Zhen Guo$^{1}$, Tian Chen$^{1,2}$}

\author{Xiang-Bin Wang$^{1,2,3}$}\email{xbwang@mail.tsinghua.edu.cn}

\affiliation{$^{1}$State Key Laboratory of Low
Dimensional Quantum Physics, Department of Physics, Tsinghua University, Beijing 100084,
People's Republic of China\\
$^2$Synergetic Innovation Center of Quantum Information and Quantum Physics, University of Science and Technology of China, Hefei, Anhui 230026, People's Republic of China\\
$^3$Jinan Institute of Quantum Technology, Shandong
Academy of Information and Communication Technology, Jinan 250101,
People's Republic of China
}

\date{\today}

\begin{abstract}
We study the non-Markovian dynamics of a damped oscillator coupled with a reservoir. We present exact formulas for the oscillator's evolution directly from the BCH formula by series expansion with neither Markovian nor rotating wave approximation (RWA). Based on these, we show the existence of the non-Markovian feature of the system's evolution for the damped oscillator.  By numerical simulation we find that the non-Markovian feature exists within a wide range of the coupling strength, even when the coupling strength is very small.
To this problem, prior art results have assumed RWA and the existence of non-Markovian feature was found when the system-reservoir coupling is strong enough. However, as we show,  given such a strong coupling, the original Hamiltonian without RWA is actually not physical. Therefore, our exact study here has thoroughly concluded the issue of non-Markovian feature.
\end{abstract}


\maketitle

\textit{Introduction}.---The theory of open systems plays a crucially important role in quantum mechanics and quantum information\cite{Breuer2002}. The interaction between the system and the environment is the quantum origin of the classical states\cite{Zurek2003}, relaxation and decoherence. Therefore its study can help design methods to protect the quantum state of a qubit\cite{Misra1977,Vitali1999,Viola2005,Khodjasteh2007,Kofman2004,Wu2002}. Moreover, interaction with the environment can be adopted to produce entanglement between several separated systems\cite{Braun2002,Paz2008}.

A damped harmonic oscillator is a frequently used elementary model in quantum open system theory\cite{Ford1965,Ford1987,Caldeira1983,Feynman1963}. Different methods such as the master equation, the Langevin equation or the path integral approach have been used to study the evolution of the damped oscillator\cite{Louisell1973,Gardiner1991,Scully1997,Feynman1963}, and usually Markovian approximation and rotating-wave approximation (RWA) are assumed. The problem is found to be accurately solvable for some specific spectra of the reservoir under these approximations and some other additional approximations\cite{Liu2001}. However, the validity of these approximations in different conditions is still not clear, and a general while convenient treatment of the problem is absent.

Non-Markovian dynamics plays an important role in quantum open systems where the backflow of information from the environment to the system is significant and therefore the Markovian approximation is not valid\cite{PhysRevLett.103.210401,ma2014crossover,PhysRevLett.99.160502,PhysRevA.60.91,PhysRevLett.82.1801,PhysRevA.55.R2531}. However, exact study of non-Markovian is numerically difficult. In Ref.~\cite{Zhang2012} Zhang et al treated the problem without the Markovian approximation through using the brilliant idea of connecting the exact master equations with the nonequilibirum Green¡¯s functions. This is a big step towards the study of the non-Markovian dynamics of quantum open systems. With their non-Markovian results, they show the non-Markovian feature of the damped oscillator's evolution if the system-bath coupling strength is larger than 0.3 for sub-Ohmic bath, but they used the RWA. Therefore, given their result it's still unclear yet whether the non-Markovian feature still exists when RWA is not assumed. First, in the original Hamiltonian without RWA one cannot blindly assume the coupling strength to be too large, otherwise the Hamiltonian is not physical. As we will show in this paper, the coupling strength upper bound is $\frac{1}{{4\sqrt \pi }}$ for sub-Ohmic bath which is significant smaller than 0.3. Second, without RWA whether there still exists the Non-Markovian dynamic feature given a value of coupling strength which physically exists. Note that the RWA is not always valid in many situations\cite{niemczyk2010circuit}, especially in the cases of strong coupling\cite{PhysRevLett.105.263603}, where the anti-rotating terms can contribute significantly to the dynamics. In this paper we study this problem, and we show the non-Markovian feature exists within a wide range of the coupling strength, even though the coupling strength is very small.

To investigate the non-Markovian dynamics of the system without RWA, here we propose a new method to study the damped harmonic oscillator by series expansion with respect to the coupling strength. The paper is arranged in this way: First we give the explicit expressions of the expansion terms and show how to apply our method to numerical calculation efficiently. After that we examine the non-Markovian dynamics of the system using our formula and show the existence of an upper bound for the coupling, below which the effect of non-Markovian dynamics seems to always appear. Then we give our deduction of the formula to second order and the series expansion expressions for each order. Finally we give our conclusions.

\textit{Series expansion formula}.---The total Hamiltonian we consider is
\begin{equation}
\label{eq:H}
H = \omega_0 a^\dag a + \sum_k \omega_k b_k^\dag b_k + \sum_k g_k (a + a^\dag)(b_k + b_k^\dag)
\end{equation}
where we have choosen $\hbar$ to be 1, $a$ and $a^\dag$ are the annihilation and creation operator of the oscillator and $b_k$ and $b_k^\dag$ are those for the $k$-mode of the reservoir, $\omega_0$, $\omega_k$ and $g_k$ are all real numbers. Here we use the position-position coupling as an example, while it's not difficult to extend to other types of coupling. And the summation in Eq.~(\ref{eq:H}) can be replaced by integration in the continuous case.

According to BCH formula, the evolution of the operator $a$ in Heisenberg picture can be represented as
$a(t) = \e^{\mi H t} a \e^{-\mi H t} = a + \mi t [H, a] - \frac{1}{2} t^2 [H, [H, a]] + \cdots$. Collecting all the terms according to the power of $g_k$ and we can get the general expression for each order of expansion. Here we show the results to the second order, and a complete expression can be found in the following.
\begin{subequations}
\label{eq:series_whole}
\begin{align}
a(t) =& \e^{-\mi \omega_0 t}a + \sum_k \bigg\{ \frac{g_k}{\omega_0 - \omega_k}(\e^{-\mi \omega_0 t} - \e^{-\mi \omega_k t}) b_k  + \frac{g_k}{\omega_0 + \omega_k}
(\e^{-\mi \omega_0 t} - \e^{\mi \omega_k t})b_k^\dag  \nonumber\\
 & - \frac{2 \mi g_k^2}{(\omega_0^2 - \omega_k^2)^2} \big[(\omega_0 + \omega_k)^2 \sin \omega_k t   + (\omega_0^2 - \omega_k^2)\omega_k t \e^{-\mi \omega_0 t} +  2\mi \omega_0 \omega_k (\e^{\mi \omega_k t} - \e^{-\mi \omega_0 t}) \big] a \nonumber\\
 & + \frac{2\mi g_k^2}{\omega _0 (\omega_0^2 - \omega_k^2)}(\omega_0 \sin \omega_k t - \omega_k \sin \omega_0 t) a^\dag \bigg\} + \cdots
 \label{eq:series_a}
\end{align}
\begin{align}
b_k(t) =& \e^{-\mi \omega_k t} b_k
	+ \frac{g_k}{\omega_k - \omega_0}(\e^{-\mi \omega_k t}
	- \e^{-\mi \omega_0 t}) a  + \frac{g_k}{\omega_k + \omega_0}(\e^{-\mi \omega_k t}
 	- \e^{\mi \omega_0 t}) a^\dag \nonumber\\
 & + \sum_{k'} \bigg\{ - \frac{2 g_k g_{k'}}{(\omega_0^2 - \omega_k^2)(\omega_k - \omega_{k'})(\omega_{k'}^2 - \omega_0^2)} \big[- \mi (\omega_k - \omega_{k'})(\omega_0^2 + \omega_k \omega_{k'}) \sin \omega_0 t \nonumber\\
 & + \omega_0 (\omega_k^2 - \omega_{k'}^2) \cos \omega_0 t + \omega_0(\omega_0^2 - \omega_k^2) \e^{-\mi\omega_{k'}t} - \omega_0(\omega_0 ^2  -  \omega_{k'}^2) \e^{-\mi \omega_k t}\big] b_{k'}\nonumber\\
 & + \frac{2 g_k g_{k'}}{(\omega_0^2 - \omega_k^2)(\omega_k + \omega_{k'})(\omega_{k'}^2 - \omega_0^2)} \big[ -\mi (\omega_k + \omega_{k'})(\omega_k \omega_{k'} - \omega_0^2) \sin \omega_0 t \nonumber\\
 & + \omega_0 (\omega_{k'}^2 - \omega_k ^2) \cos \omega_0 t + \omega_0 (\omega_k^2 - \omega_0^2) \e^{\mi \omega_{k'} t} + \omega_0 (\omega_0^2 - \omega_{k'}^2) \e^{-\mi \omega_k t} \big] b_{k'}^\dag \bigg\} +\cdots
 \label{eq:series_b}
\end{align}
\end{subequations}
And the expressions for $a^\dag(t)$ and $b_k^\dag(t)$ is just the Hermitian conjugate of Eq.~(\ref{eq:series_a}) and Eq.~(\ref{eq:series_b}).

In principle we can get the operators at any time $t$ by taking all orders of the expansion terms. While in practice it will be more efficient if we keep to the second order terms and calculate the coefficients iteratively. For example, suppose we are to calculate the evolution of $a(t)$ up to a large time $t$. We can divide the time into $N$ parts and compute the evolution in each interval successively. Define $\Delta t = t / N$ and if $g_k \Delta t \ll 1, \forall k$, it is enough to keep to the second order.

The evolution under RWA can be calculated in a similar way. Note that in this case the evolution of $a$ will only produce $a$ and $b_k$ and there is no $a^\dag$ and $b_k^\dag$ terms. So the deduction will become easier. Our results are as follows:
\begin{subequations}
\label{eq:RWA_series_whole}
\begin{align}
a(t) =& \e^{-\mi \omega_0 t}a + \sum_k \frac{g_k}{\omega_k - \omega_0} (\e^{-\mi \omega_k t} - \e^{-\mi \omega_0 t}) b_k\nonumber\\
 & + \sum_k \frac{g_k^2}{(\omega_k - \omega_0)^2}  \big[\e^{-\mi \omega_k t} - (1 - i \omega_k t + i \omega_0 t) \e^{-\mi \omega_0 t}\big]a + \cdots
 \label{eq:RWA_series_a}
\end{align}
\begin{align}
b_k(t) =& \e^{-\mi \omega_k t} b_k
	+ \frac{g_k}{\omega_k - \omega_0}(\e^{-\mi \omega_k t}
	- \e^{-\mi \omega_0 t})a \nonumber\\
 & + \sum_{k'} \frac{g_k g_{k'}}{\omega_k (\omega_{k'} - \omega_0)} \bigg(\frac{\omega_k \e^{-\mi \omega_{k'} t} - \omega_{k'} \e^{-\mi \omega_k t}}{\omega_{k'} - \omega_k}  -  \frac{\omega_k \e^{-\mi \omega_0 t} - \omega_{k'} \e^{-\mi \omega_k t}}{\omega_0 - \omega_k}\bigg)b_{k'} + \cdots
 \label{eq:RWA_series_b}
\end{align}
\end{subequations}

\begin{figure}[htbp]
\includegraphics[width = 0.9\linewidth]{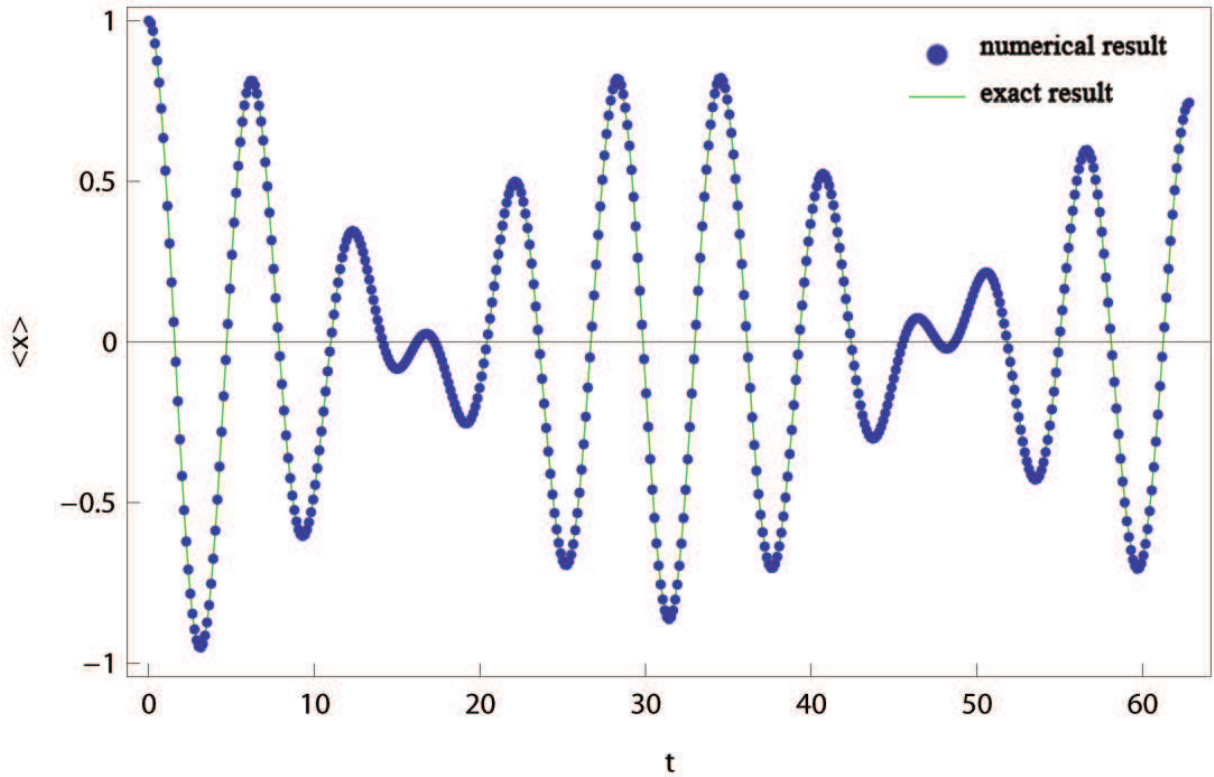}
\caption{\label{fig:Lorentz}The average position of the oscillator $\langle x \rangle$ as a function of time t. Here we choose $\Omega = 1$, $\Gamma = 0.01$ and $M = 1$, which are all defined in \cite{Liu2001}. The oscillator is initially at a coherent state $\left|\alpha\right>$ with $\alpha = 1$. The green solid line is the accurate solution and the blue points are the numerical result of our formula.}
\end{figure}

As a check of the validity of our method, we present here in Fig.~\ref{fig:Lorentz} our results for a bath of Lorentzian spectrum with RWA, which is exactly solvable\cite{Liu2001}.

Note that here we use the Lorentz spectrum only to show the correctness of our formula. In what follows we shall study the non-Markovian dynamics of Ohmic and sub-Ohmic bath.

\begin{figure}[htbp]
\includegraphics[width=\linewidth]{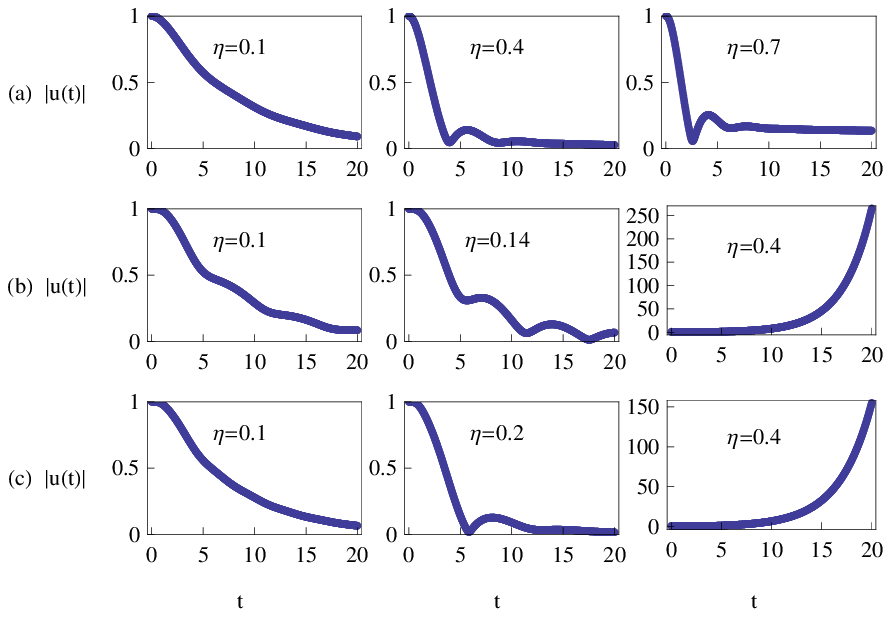}
\caption{\label{fig:Markovian}Absolute value of Green's function $u(t)$ for damped oscillator. In all these figures, we set $\omega_0 = 1$ and $\omega_c = 1$ for the Ohmic or sub-Ohmic spectrum($s = 0.5$) which are defined in \cite{Zhang2012}. (a)Results for sub-Ohmic spectrum under RWA with $\eta = 0.1,\, 0.4,\, 0.7$ respectively. We see they are consistent with those in \cite{Zhang2012}. (b)Results for sub-Ohmic spectrum without RWA, $\eta = 0.1,\, 0.14,\, 0.4$, respectively. (c)Results for Ohmic spectrum without RWA, $\eta = 0.1,\, 0.2,\, 0.4$, respectively. Obviously $|u(t)|$ oscillates easier with the existence of the anti-rotating term, and when $\eta$ is too large the figure increases exponentially. This means the total Hamiltonian is unphysical.}
\end{figure}

\textit{Non-Markovian dynamics}.---Now we use our method to study the non-Markovian dynamics of the oscillator. We should consider a general non-
Markovian environment with spectral density
\begin{equation}
J(\omega ) = 2\pi \eta \omega {(\frac{\omega }{{{\omega _c}}})^{s - 1}}\exp ( - \frac{\omega }{{{\omega _c}}})
\end{equation}
where $\eta$ is a constant describing the coupling strength between the system and the environment, and ${\omega _c}$ is the frequency cutoff. When $s = 1$, $<1$ and $>1$, the corresponding environments are Ohmic, sub-Ohmic and super-Ohmic, respectively. The non-Markovian dynamics of the oscillator can be characterized by a function $u(t) = [a(t), a^\dag(0)]$\cite{Zhang2012}. In Fig.~\ref{fig:Markovian} we repeat the results under RWA in \cite{Zhang2012} and compare it with those without RWA. Obviously the coupling strength plays a crucially important role in non-Markovian dynamics. The larger the coupling, the more significant the non-Markovian effects. Also we can see that with the existence of the anti-rotating terms, the non-exponential decay becomes more important and the curve of $|u(t)|$ begins to oscillate at smaller coupling $\eta$.

However, the coupling of a real physical system cannot be too large. Note that in Fig.~\ref{fig:Markovian}, $|u(t)|$ for non-RWA cases increases exponentially when $\eta$ is large, which is surely not physical. This is different from the RWA Hamiltonian, for which the evolution of the system seems to be always bounded given whatever strong coupling strength. Actually, as we shall see below, there is an upper bound for the coupling between the oscillator and the environment so as to make the model meaningful where the reservoir is represented by a series of independent oscillators coupled with the original oscillator.

As is in Eq.~(\ref{eq:H}), we begin with a discrete spectrum. Take the position-position coupling as an example.  We rewrite the Hamiltonian in this way:
 \begin{equation}
  H = \frac 12 P^\dagger P + \frac 12 X^\dagger V X
 \end{equation}
 where $P = (p_a, p_{b_1} , p_{b_2}, \cdots )^{\mathrm T}$, $X = ( x_a, x_{b_1}, x_{b_2}, \cdots )^{\mathrm T}$, $x$ ($p$) represents the position (momentum) operator. For simplicity, we set $m = 1$. The first term
 is the kinetic energy term, which is already diagonal, and the second term is potential term, where $V$ is:
\begin{equation}
V =   \begin{pmatrix}
\omega_0^2 & 2 g_1 \sqrt{\omega_0 \omega_1} & \ldots & 2 g_k \sqrt{\omega_0 \omega_k} & \ldots\\
2 g_1 \sqrt{\omega_0 \omega_1} & \omega_1^2 & & 0 &\\
\vdots & & \ddots &\\
2 g_k \sqrt{\omega_0 \omega_k} & 0 & & \omega_k^2 &\\
\vdots & & & & \ddots
\end{pmatrix}
\end{equation}

The eigenvalues of the matrix $V$, which correspond to the square of the frequency of each normal mode of the whole system, must be non-negative. Otherwise the potential of the system won't have a lower limit and the evolution of the oscillator will rise exponentially, which is exactly the case in Fig.~\ref{fig:Markovian}.

To get the upper bound of the coupling, we consider the critical condition where one of the eigenvalue of $V$ is 0, which leads to
\begin{equation}
\label{eq:CritDis}
4 \sum_k \frac{g_k^2}{\omega_0 \omega_k} = 1
\end{equation}

\begin{figure}[htbp]
\begin{center}
\includegraphics[width=0.8\linewidth]{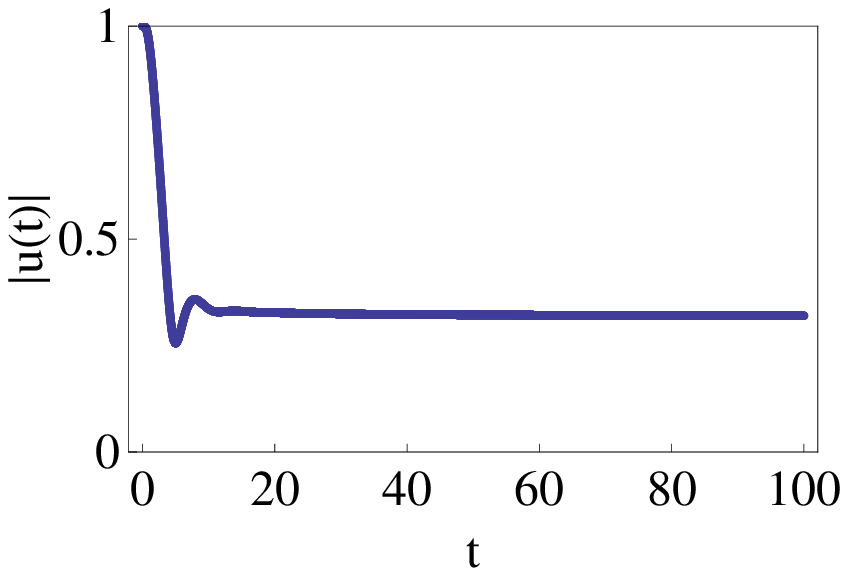}
\end{center}
\caption{\label{fig:crit}Absolute value of Green's function $u(t)$ for an Ohmic spectrum with $\omega_0 = \omega_c =1$, $\eta = 0.25$. Note that in this critical condition $|u(t)|$ keeps constant as $t$ increases.}
\end{figure}

Eq.~(\ref{eq:CritDis}) can be generalized to the case of a continuous spectrum simply by replacing the summation $\sum_k g_k^2$ with integration $\int{\mathrm{d}}\omega J(\omega )/2\pi$. Take the standard ohmic bath as an example. We get the upper bound $\eta_M = \omega_0 / 4\omega_c$ for the potential to have a lower limit. In the case of $\omega_0 = \omega_c =1$ it gives $\eta_M = 0.25$, which is consistent with our numerical results (see Fig.~\ref{fig:crit}). And for sub-Ohmic bath where we choose $s = 0.5$, it's easy to obtain $\eta_M = \frac{1}{{4\sqrt \pi  }} \approx 0.141 < 0.3$.

The above results show that, even though in the case of RWA large $\eta$ still gives bounded evolution, it does not mean the value of $\eta$ can be taken at will. The value of $\eta$ should not exceed $\eta_M$ given above, for otherwise the RWA is physically meaningless.
And simple calculation will show that any non-zero value of coupling will lead to unphysical results for the Lorentz spectrum because the integration (in place of the summation) in Eq.~(\ref{eq:CritDis}) does not converge when the frequency is small.

\begin{figure}
\includegraphics[width=1\linewidth]{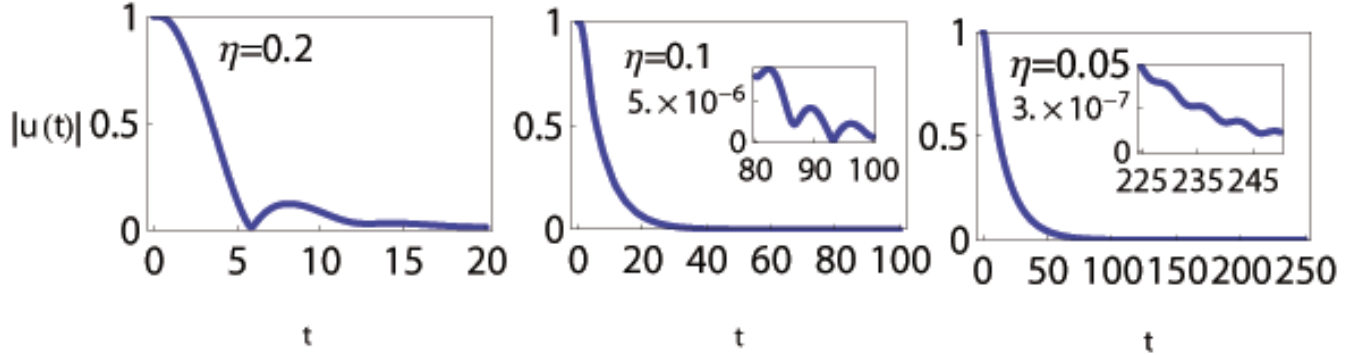}
\caption{\label{fig:Markovian_Ohm}Absolute value of the Green's function $u(t)$ for an Ohmic spectrum with $\omega_0 = \omega_c =1$, $\eta = 0.2,\, 0.1,\, 0.05$, respectively. As $\eta$ decreases, the starting time of oscillation increases and the amplitude of the oscillation decreases. So when $\eta$ is sufficiently small, the non-Markovian property of the system is not significant.}
\end{figure}

Therefore, to study the non-Markovian dynamics of the system, RWA with too strong coupling is not convincing such as $\eta = 0.3$ for sub-Ohmic bath. We should examine the non-RWA solution with coupling strength less than the upper limit given by Eq.~(\ref{eq:CritDis}). In the case of an Ohmic spectrum $\omega_0 = \omega_c =1$, we present in Fig.~\ref{fig:Markovian_Ohm} the $|u(t)|-t$ curves for several different $\eta$, all of which oscillate and therefore show significant non-Markovian dynamics. And as we can see, when $\eta$ decreases, the time when $|u(t)|$ begins to oscillate increases and the amplitude of such oscillation decreases. So we have firmly proven the existence of the non-Markovian property of the system. And it seems that any non-zero $\eta$ can produce such effects. Moreover, we've shown that when the coupling is sufficiently small, the backflow of information from the environment is negligible and therefore the Markovian approximation is viable.

\textit{Deduction}.---In this part we show the derivation of Eqs.~(\ref{eq:series_whole}). First for simplify we redefine the operators $\mathcal{X} = a^\dag + a$, $\mathcal{P} = a^\dag - a$, and define $T_0 = \mathcal X_a$,$T_n = [H,T_{n - 1}]$, where $H$ is the Hamiltonian defined by Eq.~(\ref{eq:H}). Then BCH formula gives $\mathcal X_a(t) = \e^{\mi Ht} \mathcal X_a \e^{- \mi Ht} = \sum_{n = 0}^\infty \frac{( \mi t)^n}{n!} T_n$. 

The basic commutation relationships we need are:
\begin{subequations}
\begin{align}
[H,\mathcal X_a] &= \omega_0 \mathcal P_a\\
[H,\mathcal P_a] &= \omega_0 \mathcal X_a + 2\sum_k g_k \mathcal X_{bk}\\
[H,\mathcal X_{bk}] &= \omega_k \mathcal P_{bk}\\
[H,\mathcal P_{bk}] &= \omega_k \mathcal X_{bk} + 2 g_k \mathcal X_a
\end{align}
\end{subequations}

With these formulae we can get the recurrence relations for $n \ge 1$:
\begin{subequations}
\label{eq:recurrence}
\begin{align}
\begin{aligned}
T_{2n} =& \omega_0^2 T_{2n-2} + 4 \omega_0 \sum_{i=1}^{n-1} \sum_k g_k^2 \omega_k^{2i-1} T_{2n-2i-2} \\
& + 2 \omega_0 \sum_k g_k \omega_k^{2n-2} \mathcal X_{bk}
\end{aligned}\label{eq:even}\\
\begin{aligned}
T_{2n+1} =& \omega_0^2 T_{2n-1} + 4 \omega_0 \sum_{i=1}^{n-1} \sum_k g_k^2 \omega_k^{2i-1} T_{2n-2i-1} \\
& + 2 \omega_0 \sum_k g_k \omega_k^{2n-1} \mathcal P_{bk}
\end{aligned}\label{eq:odd}
\end{align}
\end{subequations}
Note that $T_{2n}$ contains only terms about $\mathcal X$ while $T_{2n+1}$ only $\mathcal P$.

Now we define $T_{n,i}$ as the $i^{\mathrm {th}}$ order term in ${T_n}$ with respect to the coupling strength. Obviously, $T_{2n,0} = \omega_0^{2n} \mathcal X_a$. So from Eq.~(\ref{eq:even}) we can get
\begin{subequations}
\begin{align}
T_{2n,1} &= \omega_0^2 T_{2n-2,1} + 2 \omega_0 \sum_k g_k \omega_k^{2n-2} \mathcal X_{bk}\\
T_{2n,2} &= \omega_0^2 T_{2n-2,2} + 4 \omega_0 \sum_{i=1}^{n-1} \sum_k g_k^2 \omega_k^{2i-1} \omega_0^{2n-2i-2} \mathcal X_a
\end{align}
\end{subequations}
for $n \ge 1$.

Since $T_{2,2} = 0$, $T_{2,1} = 2 \omega_0 \sum_k g_k \mathcal X_{bk}$, finally we obtain
\begin{subequations}
\begin{align}
T_{2n,1} &= \sum_{r = 0}^{n - 1} 2 \omega_0^{2r + 1} \sum_k g_k \omega_k^{2n - 2 - 2r} \mathcal X_{bk} \nonumber\\
 &=2 \sum_k g_k \omega_0 \left( \frac{\omega_0^{2n} - \omega_k^{2n}}{\omega_0^2 - \omega_k^2} \right) \mathcal X_{bk} \\
T_{2n,2} &= 4\sum_{r = 1}^n \sum_{i = 1}^{n - r} \sum_k {g_k^2} \omega_k^{2i - 1}\omega_0^{2n - 2i - 1} \mathcal X_a \nonumber\\
 &= 4\sum_k g_k^2 \frac{\omega_0^{2n - 3} \omega_k}{\left( \frac{\omega_k}{\omega_0} \right)^2 - 1} \left[ \frac{\left( \frac{\omega_k}{\omega_0} \right)^{2n} - 1}{\left( \frac{\omega_k}{\omega_0} \right)^2 - 1} - n \right] \mathcal X_a
\end{align}
\end{subequations}

We can get the results about $T_{2n + 1}$ in a similar way:
\begin{subequations}
\begin{align}
T_{2n + 1,1} &= 2 \sum_k g_k \omega_k \omega_0 \left( \frac{\omega_0^{2n} - \omega_k^{2n}}{\omega_0^2 - \omega_k^2} \right) \mathcal P_{bk}\\
T_{2n + 1,2} &= 4\sum_k g_k^2 \frac{\omega_0^{2n - 2}\omega_k}{\left( \frac{\omega_k}{\omega_0} \right)^2 - 1}\left[ \frac{\left( \frac{\omega_k}{\omega_0} \right)^{2n} - 1}{\left( \frac{\omega_k}{\omega_0} \right)^2 - 1} - n \right] \mathcal P_a
\end{align}
\end{subequations}

Eventually, we get the second order expression:
\begin{align}
  \mathcal X_a(t) =& \sum_{n=0}^\infty \frac{(\mi t)^n}{n!} T_n = \sum_n \frac{(\mi t)^n}{n!}(T_{n,0} + T_{n,1} + T_{n,2} + \cdots) \nonumber \\
   =& a^\dag \e^{\mi \omega_0 t} + a \e^{-\mi \omega_0 t} \nonumber \\
   & + \sum_k \bigg\{\frac{2 g_k \omega_0 \mathcal X_{bk}}{\omega_0^2 - \omega_k^2}(\cos \omega_0 t - \cos \omega_k t) +  \frac{2\mi g_k \mathcal P_{bk}}{\omega_0^2 - \omega_k^2}(\omega_k \sin \omega_0 t - \omega_0 \sin \omega_k t) \nonumber \\
   & + \frac{4 g_k^2 \omega_k \mathcal X_a}{(\omega_k^2 - \omega_0^2)^2}  \left[ \omega_0 (\cos \omega_k t - \cos \omega_0 t) + \frac{1}{2}(\omega_k^2 - \omega_0^2) t \sin \omega_0 t \right] \nonumber \\
   & + \frac{4\mi g_k^2 \mathcal P_a}{(\omega_k^2 - \omega_0^2)^2} \bigg[ \omega_0^2 \sin \omega_k t - \omega_0 \omega_k \sin \omega_0 t  - \frac{\omega_k (\omega_k^2 - \omega_0^2)}{\omega_0} \frac{\omega_0 t \cos \omega_0 t - \sin \omega_0 t}{2} \bigg] \bigg\} + \cdots
\end{align}

It is easy to obtain the expressions of $\mathcal P_a(t)$, $\mathcal X_{bk}(t)$ and $\mathcal P_{bk}(t)$ in the same way. Therefore we can get Eqs.~(\ref{eq:series_whole}). And from Eqs.~(\ref{eq:recurrence}) we can get the expressions for higher order terms in a similar way.

For RWA, we can define $T'_0 = a$ and $T'_n = [H_R,T'_{n-1}]$, where $H_R = \omega_0 a^\dag a + \sum_k \omega_k b_k^\dag b_k + \sum_k g_k (a b_k^\dag + a^\dag b_k)$ is the Hamiltonian under RWA. Using the recurrence relation:
\begin{equation}
T'_n = - \omega_0 T'_{n-1} + \sum_{i=0}^{n-2} \sum_k g_k^2 (- \omega_k)^i T'_{n-2-i} - \sum_k g_k (-\omega_k)^{n-1} b_k
\end{equation}
we can get the expressions for RWA.

\textit{Series expansion expressions for each order}.---Here we give the expressions for series expansion to each order. Consider a more general system with several independent oscillators instead of only one. The Hamiltonian of the system is:
\begin{equation}
H = \sum_i \Omega_i a_i^\dag a_i + \sum_k \omega_k b_k^\dag b_k + \sum_i \sum_k g_{ik} (a_i^\dag + a_i)(b_k^\dag + b_k)
\end{equation}
where we also have $\hbar=1$. $\Omega_i$, $a_i$ and $a_i^\dag$ are the frequency, annihilation and creation operator of the ${i^{th}}$ oscillator, respectively, while $\omega_k$,$b_k$ and $b_k^\dag$ are those for the k-mode of the reservoir. And $g_{ik}$ is the coupling strength between the ${i^{th}}$ oscillator and the k-mode of the reservoir.

Under this condition, the complete expressions for series expansion with respect to the coupling strength are:
\begin{subequations}
\begin{align}
  \mathcal{X}_{a i_1}(t) =& a_{i_1}\e^{-\mi \Omega_{i_1}t} + a_{i_1}^\dag \e^{\mi \Omega _{i_1}t} \nonumber\\
  & + \sum_{n=1}^\infty \sum_{i_2} \sum_{i_3} \cdots  \sum_{i_n} \sum_{k_1} \sum_{k_2} \cdots \sum_{k_n} 2^{2n-1} g_{i_1 k_1} \prod_{r=2}^n g_{i_r k_{r-1}} g_{i_r k_r} \prod_{m=1}^{n-1} \Omega_{i_m} \omega_{k_m} \nonumber\\
  &\qquad \cdot \Omega_{i_n} \sum_{l=1}^n \bigg[ \frac{\cos \Omega_{i_l} t}{(\Omega_{i_l}^2 - \omega_{k_l}^2) \prod_{l'=1,l' \ne l}^n (\Omega_{i_l}^2 - \Omega_{i_{l'}}^2)(\Omega_{i_l}^2 - \omega_{k_{l'}}^2)} \nonumber\\
  &\qquad \qquad \qquad + \frac{\cos \omega_{k_l}t}{(\omega_{k_l}^2 - \Omega_{i_l}^2) \prod_{l'=1,l' \ne l}^n (\omega_{k_l}^2 - \Omega_{i_{l'}}^2)(\omega_{k_l}^2 - \omega_{k_{l'}}^2)}\bigg] \mathcal{X}_{b k_n} \nonumber \\
   & + \sum_{n=1}^\infty \sum_{i_2} \sum_{i_3} \cdots \sum_{i_{n+1}} \sum_{k_1} \sum_{k_2} \cdots \sum_{k_n} 2^{2n} g_{i_1 k_1} \prod_{r=2}^n g_{i_r k_{r-1}} g_{i_r k_r} \cdot g_{i_{n+1} k_n} \prod_{m=1}^n \Omega_{i_m} \omega_{k_m} \nonumber \\
  & \qquad \cdot \Bigg[ \sum_{l=1}^{n+1} \frac{\cos \Omega_{i_l} t}{\prod_{l'=1,l' \ne l}^{n+1} (\Omega_{i_l}^2 - \Omega_{i_{l'}}^2) \prod_{l'=1}^{n} (\Omega_{i_l}^2 - \omega_{k_{l'}}^2)}  \nonumber \\
  & \qquad \qquad \qquad + \sum_{l=1}^n \frac{\cos \omega_{k_l} t}{\prod_{l'=1}^{n+1} (\omega_{k_l}^2 - \Omega_{i_{l'}}^2) \prod_{l'=1,l' \ne l}^n (\omega_{k_l}^2 - \omega_{k_{l'}}^2)} \Bigg] \mathcal{X}_{a i_{n+1}} \nonumber \\
  & + \mi \sum_{n=1}^\infty \sum_{i_2} \sum_{i_3} \cdots \sum_{i_n} \sum_{k_1} \sum_{k_2} \cdots \sum_{k_n} 2^{2n - 1} g_{i_1 k_1} \prod_{r=2}^n g_{i_r k_{r-1}} g_{i_r k_r} \prod_{m=1}^n \Omega_{i_m} \omega_{k_m} \nonumber \\
  & \qquad \cdot \sum_{l=1}^n \bigg[ \frac{\sin \Omega_{i_l} t}{\Omega_{i_l} (\Omega_{i_l}^2 - \omega_{k_l}^2) \prod_{l'=1,l' \ne l}^n (\Omega_{i_l}^2 - \Omega_{i_{l'}}^2)(\Omega_{i_l}^2 - \omega_{k_{l'}}^2)} \nonumber \\
  & \qquad \qquad \qquad +\frac{\sin \omega_{k_l} t}{\omega_{k_l} (\omega_{k_l}^2 - \Omega_{i_l}^2) \prod_{l'=1,l' \ne l}^n (\omega_{k_l}^2 - \Omega_{i_{l'}}^2)(\omega_{k_l}^2 - \omega_{k_{l'}}^2) } \bigg] \mathcal{P}_{b k_n} \nonumber \\
  & + \mi \sum_{n=1}^\infty \sum_{i_2} \sum_{i_3} \cdots \sum_{i_{n+1}} \sum_{k_1} \sum_{k_2} \cdots \sum_{k_n} 2^{2n} g_{i_1 k_1} \prod_{r=2}^n g_{i_r k_{r-1}} g_{i_r k_r} \cdot g_{i_{n+1} k_n} \prod_{m=1}^n \Omega_{i_m} \omega_{k_m} \nonumber \\
  & \qquad \cdot \Omega_{i_{n+1}} \Bigg[ \sum_{l=1}^{n+1} \frac{\sin \Omega_{i_l} t}{\Omega_{i_l} \prod_{l'=1,l' \ne l}^{n+1} (\Omega_{i_l}^2 - \Omega_{i_{l'}}^2) \prod_{l'=1}^n (\Omega_{i_l}^2 - \omega_{k_{l'}}^2)} \nonumber \\
  & \qquad \qquad \qquad + \sum_{l=1}^n \frac{\sin \omega_{k_l} t}{\omega_{k_l} \prod_{l'=1}^{n+1} (\omega_{k_l}^2 - \Omega_{i_{l'}}^2) \prod_{l'=1,l' \ne l}^n (\omega_{k_l}^2 - \omega_{k_{l'}}^2)} \Bigg] \mathcal{P}_{a{i_{n+1}}}
\end{align}
\begin{align}
\mathcal{P}_{a i_1}(t) =& -a_{i_1}\e^{-\mi \Omega_{i_1}t} + a_{i_1}^\dag \e^{\mi \Omega _{i_1}t} \nonumber\\
  & + \sum_{n=1}^\infty \sum_{i_2} \sum_{i_3} \cdots  \sum_{i_n} \sum_{k_1} \sum_{k_2} \cdots \sum_{k_n} 2^{2n-1} g_{i_1 k_1} \prod_{r=2}^n g_{i_r k_{r-1}} g_{i_r k_r} \prod_{m=2}^n \Omega_{i_m} \omega_{k_m} \nonumber\\
  &\qquad \cdot \omega_{k_1} \sum_{l=1}^n \bigg[ \frac{\cos \Omega_{i_l} t}{(\Omega_{i_l}^2 - \omega_{k_l}^2) \prod_{l'=1,l' \ne l}^n (\Omega_{i_l}^2 - \Omega_{i_{l'}}^2)(\Omega_{i_l}^2 - \omega_{k_{l'}}^2)} \nonumber\\
  &\qquad \qquad \qquad + \frac{\cos \omega_{k_l}t}{(\omega_{k_l}^2 - \Omega_{i_l}^2) \prod_{l'=1,l' \ne l}^n (\omega_{k_l}^2 - \Omega_{i_{l'}}^2)(\omega_{k_l}^2 - \omega_{k_{l'}}^2)}\bigg] \mathcal{P}_{b k_n} \nonumber \\
   & + \sum_{n=1}^\infty \sum_{i_2} \sum_{i_3} \cdots \sum_{i_{n+1}} \sum_{k_1} \sum_{k_2} \cdots \sum_{k_n} 2^{2n} g_{i_1 k_1} \prod_{r=2}^n g_{i_r k_{r-1}} g_{i_r k_r} \cdot g_{i_{n+1} k_n} \prod_{m=1}^n \Omega_{i_{m+1}} \omega_{k_m} \nonumber \\
  & \qquad \cdot \Bigg[ \sum_{l=1}^{n+1} \frac{\cos \Omega_{i_l} t}{\prod_{l'=1,l' \ne l}^{n+1} (\Omega_{i_l}^2 - \Omega_{i_{l'}}^2) \prod_{l'=1}^{n} (\Omega_{i_l}^2 - \omega_{k_{l'}}^2)}  \nonumber \\
  & \qquad \qquad \qquad + \sum_{l=1}^n \frac{\cos \omega_{k_l} t}{\prod_{l'=1}^{n+1} (\omega_{k_l}^2 - \Omega_{i_{l'}}^2) \prod_{l'=1,l' \ne l}^n (\omega_{k_l}^2 - \omega_{k_{l'}}^2)} \Bigg] \mathcal{P}_{a i_{n+1}} \nonumber \\
  & + \mi \sum_{n=1}^\infty \sum_{i_2} \sum_{i_3} \cdots \sum_{i_n} \sum_{k_1} \sum_{k_2} \cdots \sum_{k_n} 2^{2n - 1} g_{i_1 k_1} \prod_{r=2}^n g_{i_r k_{r-1}} g_{i_r k_r} \prod_{m=1}^{n-1} \Omega_{i_{m+1}} \omega_{k_m} \nonumber \\
  & \qquad \cdot \sum_{l=1}^n \bigg[ \frac{\Omega_{i_l} \sin \Omega_{i_l} t}{(\Omega_{i_l}^2 - \omega_{k_l}^2) \prod_{l'=1,l' \ne l}^n (\Omega_{i_l}^2 - \Omega_{i_{l'}}^2)(\Omega_{i_l}^2 - \omega_{k_{l'}}^2)} \nonumber \\
  & \qquad \qquad \qquad +\frac{\omega_{k_l} \sin \omega_{k_l} t}{(\omega_{k_l}^2 - \Omega_{i_l}^2) \prod_{l'=1,l' \ne l}^n (\omega_{k_l}^2 - \Omega_{i_{l'}}^2)(\omega_{k_l}^2 - \omega_{k_{l'}}^2) } \bigg] \mathcal{X}_{b k_n} \nonumber \\
  & + \mi \sum_{n=1}^\infty \sum_{i_2} \sum_{i_3} \cdots \sum_{i_{n+1}} \sum_{k_1} \sum_{k_2} \cdots \sum_{k_n} 2^{2n} g_{i_1 k_1} \prod_{r=2}^n g_{i_r k_{r-1}} g_{i_r k_r} \cdot g_{i_{n+1} k_n} \prod_{m=2}^n \Omega_{i_m} \omega_{k_m} \nonumber \\
  & \qquad \cdot \omega_{k_1} \Bigg[ \sum_{l=1}^{n+1} \frac{\Omega_{i_l} \sin \Omega_{i_l} t}{\prod_{l'=1,l' \ne l}^{n+1} (\Omega_{i_l}^2 - \Omega_{i_{l'}}^2) \prod_{l'=1}^n (\Omega_{i_l}^2 - \omega_{k_{l'}}^2)} \nonumber \\
  & \qquad \qquad \qquad + \sum_{l=1}^n \frac{\omega_{k_l} \sin \omega_{k_l} t}{\prod_{l'=1}^{n+1} (\omega_{k_l}^2 - \Omega_{i_{l'}}^2) \prod_{l'=1,l' \ne l}^n (\omega_{k_l}^2 - \omega_{k_{l'}}^2)} \Bigg] \mathcal{X}_{a{i_{n+1}}}
\end{align}
\end{subequations}
where we define $\mathcal{X} = a^\dag + a$, $\mathcal{P} = a^\dag - a$. 

Under RWA, the expressions are much simpler:
\begin{align}
  a_{i_1}(t) =& a_{i_1}\e^{-\mi \Omega_{i_1}t} + \sum_{n = 1}^\infty \sum_{i_2} \sum_{i_3} \cdots \sum_{i_n} \sum_{k_1} \sum_{k_2} \cdots \sum_{k_n} g_{i_1 k_1} \prod_{r=2}^n g_{i_r k_{r-1}} g_{i_r k_r} \nonumber\\
  & \qquad \cdot \sum_{l = 1}^n \bigg[ \frac{\exp (-\mi \Omega_{i_l} t)}{(\Omega_{i_l} - \omega_{k_l}) \prod_{l'=1,l' \ne l}^n (\Omega_{i_l} - \Omega_{i_{l'}})(\Omega_{i_l} - \omega_{k_{l'}})}  \nonumber\\
  & \qquad \qquad \qquad + \frac{\exp (-\mi \omega_{k_l} t)}{(\omega_{k_l} - \Omega_{i_l}) \prod_{l'=1,l' \ne l}^n (\omega_{k_l} - \Omega_{i_{l'}})(\omega_{k_l} - \omega_{k_{l'}})} \bigg] b_{k_n} \nonumber\\
   & + \sum_{n = 1}^\infty \sum_{i_2} \sum_{i_3} \cdots \sum_{i_{n+1}} \sum_{k_1} \sum_{k_2} \cdots \sum_{k_n} g_{i_1 k_1} \prod_{r=2}^n g_{i_r k_{r-1}} g_{i_r k_r}  \nonumber\\
  & \qquad \cdot \Bigg[ \sum_{l=1}^{n+1} \frac{\exp (-\mi \Omega_{i_l}t)}{\prod_{l'=1,l' \ne l}^{n+1} (\Omega_{i_l} - \Omega_{i_{l'}}) \prod_{l'=1}^n (\Omega_{i_l} - \omega_{k_{l'}})} \nonumber\\
  & \qquad \qquad \qquad + \sum_{l = 1}^n \frac{\exp (-\mi \omega_{k_l}t)}{\prod_{l'=1}^{n+1} (\omega_{k_l} - \Omega_{i_{l'}}) \prod_{l'=1,l' \ne l}^n (\omega_{k_l} - \omega_{k_{l'}})} \Bigg] a_{i_{n+1}}
\end{align}
By taking Hermitian conjugate we get the expression for $a_{i_1}^\dag (t)$. And since the terms relevant to oscillators and those to the reservoir appear symmetrically in the Hamiltonian, it's easy to get the expressions for each mode of the reservoir by simplify exchange the corresponding terms.

\textit{Conclusion}.---In this paper we present our exact formula for the motion of a damped harmonic oscillator without Markovian or rotating-wave approximation. The correctness of the formula has been proven for the examples we give, comparing with different previous works about different models of the reservoir. With this formula, we study the non-Markovian dynamics of the system and show that the backflow of information exists when including the anti-rotating interaction. We also show the existence of non-Markovian property for a wide range of coupling strength and that such effects weaken when the coupling strength decreases.

\textit{Acknowledgments}.---X. B. Wang proposed this work. M. J. Tang, Y. K. Wu and M. Lyu equally contributed to this work. We acknowledge the financial support in part by the 10000-Plan of Shandong province, the National High-Tech Program of China Grants No. 2011AA010800 and 2011AA010803, NSFC Grants No. 11174177 and 60725416.

\end{document}